\documentclass[12pt]{article}

\usepackage{scicite}

\usepackage{times}

\usepackage{graphicx}
\usepackage{bm}

\newenvironment{sciabstract}{%
\begin{quote} \bf}
{\end{quote}}

\newcounter{lastnote}

\title{Quantum Hall effect in a bulk antiferromagnet EuMnBi$_2$ with magnetically confined two-dimensional Dirac fermions} 
\author
{Hidetoshi Masuda,$^{1\dagger}$ Hideaki Sakai,$^{1,2\dagger\ast}$ Masashi Tokunaga,$^{3}$ Yuichi Yamasaki,$^{4,5}$\\ Atsushi Miyake,$^{3}$ Junichi Shiogai,$^{6}$ Shintaro Nakamura,$^{6}$ Satoshi Awaji,$^{6}$\\ Atsushi Tsukazaki,$^{6}$ Hironori Nakao,$^7$ Youichi Murakami,$^7$ Taka-hisa Arima,$^{5,8}$\\ Yoshinori Tokura,$^{1,5}$ Shintaro Ishiwata$^{1,9}$\\
\\
\normalsize{$^{1}$Department of Applied Physics, University of Tokyo, Tokyo 113-8656, Japan}\\
\normalsize{$^{2}$Department of Physics, Osaka University, Toyonaka, Osaka 560-0043, Japan}\\
\normalsize{$^{3}$The Institute for Solid State Physics, University of Tokyo, Kashiwa 277-8581, Japan}\\
\normalsize{$^{4}$Department of Applied Physics and Quantum-Phase Electronics Center (QPEC),}\\
\normalsize{University of Tokyo, Tokyo 113-8656, Japan}\\
\normalsize{$^{5}$RIKEN Center for Emergent Matter Science (CEMS), Wako 351-0198, Japan}\\
\normalsize{$^{6}$Institute for Materials Research, Tohoku University, Sendai 980-8577, Japan}\\
\normalsize{$^{7}$Condensed Matter Research Center and Photon Factory, Institute of Materials}\\
\normalsize{Structure Science, KEK, Tsukuba 305-0801, Japan}\\
\normalsize{$^{8}$Department of Advanced Materials Science, University of Tokyo, Kashiwa 277-8561, Japan}\\
\normalsize{$^{9}$PRESTO, Japan Science and Technology Agency, Kawaguchi, Saitama 332-0012, Japan.}\\
\\
\normalsize{$^\dagger$These authors contributed equally to this work.}\\
\normalsize{$^\ast$To whom correspondence should be addressed; E-mail: sakai@phys.sci.osaka-u.ac.jp.}
}

\date{}

\begin{document} 


\baselineskip24pt


\maketitle 

\begin{sciabstract}
For the innovation of spintronic technologies, Dirac materials, in which the low-energy excitation is described as relativistic Dirac fermions, are one of the most promising systems, because of the fascinating magnetotransport associated with the extremely high mobility.
To incorporate Dirac fermions into spintronic applications, their quantum transport phenomena are desired to be manipulated to a large extent by magnetic order in a solid.
We here report a bulk half-integer quantum Hall effect in a layered antiferromagnet EuMnBi$_2$, in which field-controllable Eu magnetic order significantly suppresses the interlayer coupling between the Bi layers with Dirac fermions.
In addition to the high mobility more than 10,000 cm$^2$/Vs, Landau level splittings presumably due to the lifting of spin and valley degeneracy are noticeable even in a bulk magnet.
These results will pave a route to the engineering of magnetically functionalized Dirac materials.
\end{sciabstract}
%
%
\section*{Introduction}
%
A conductive material with magnetic order is an integral component for spintronic devices, such as spin valves and spin transistors\cite{Zutic2004RMP}.
There, charge transport correlated with magnetism, such as giant magnetoresistance effect, enables high-speed and/or nonvolatile device operations.
Dirac fermions with linear energy dispersion with momentum space have been of current interest for spintronic applications, since a variety of quantum transport phenomena manifest themselves in an external magnetic field, due to the extremely high mobility.
A typical example is an unusual half-integer quantum Hall effect (QHE)\cite{Novoselov2005Nature,Zhang2005Nature} observable even at room temperature in graphene\cite{Novoselov2007Science}.
More recently, the quantum anomalous Hall effect was observed for the surface Dirac state in magnetic topological insulator thin films\cite{Chang2013Science,Checkelsky2014NatPhys}.
To expand potential application of such a distinct quantum transport enriched by magnetic order, it is highly desirable to explore bulk systems that host various magnetism and dimensionality. 
%
\par
%
Despite recent discovery of a number of new bulk Dirac materials, novel quantum transport features have been elucidated mostly in nonmagnetic materials, as exemplified by so-called Dirac/Weyl semimetals, such as Cd$_{3}$As$_2$\cite{Liang2014NatMat, He2014PRL, Narayanan2015PRL}, Na$_3$Bi\cite{Kushwaha2015APLMat}, and TaAs\cite{Huang2015Arxive}.
Since the emergence of the three-dimensional Dirac-like dispersion stems from specific lattice symmetry for the above materials, it would be in principle impossible to substitute the constituent elements with magnetic ones with keeping the crystal structure.
As strongly correlated magnetic systems, on the other hand, certain heavy transition metal oxides, such as pyrochlore irridates, have been predicted to have Weyl semimetallic states\cite{Krempa2014ARCMP}.
At present, however, quantum transport phenomena associated with Dirac-like quasiparticles remain experimentally elusive.
\par
%
In this context, layered compound $A$MnBi$_2$ [$A$=Sr$^{2+}$\cite{Park2011PRL,Wang2011PRB,K_Wang201PRB} and Eu$^{2+}$\cite{May2014PRB}] would provide an ideal arena to reveal the interplay between Dirac fermions and ordered magnetic moments.
This is because the conducting layers of Bi square net hosting quasi two-dimensional (2D) Dirac fermions and the insulating magnetic layers consisting of the Mn-Bi and $A$ layers are spatially separated (Fig. 1E), where we can develop a variety of magnetic layers while keeping the Dirac-like band structure.
For EuMnBi$_2$, importantly, a signature of coupling between charge transport and magnetism was recently discerned upon the magnetic order of Eu moments (spin $S\!=\!7/2$ for Eu$^{2+}$) adjacent to the Bi layer\cite{May2014PRB}.
By applying fields up to 55 T that enable complete control of the magnetic order of Eu sublattice, we here demonstrate its strong impact on interlayer hopping of quasi 2D Dirac fermions on the Bi layer, which gives rise to the multilayer quantum Hall state.
\section*{Results and Discussions}
%
As shown in Fig. 1A, the magnetic susceptibility $M/H$ parallel to the $c$ axis for EuMnBi$_{2}$ steeply decreases below the antiferromagnetic (AFM) transition temperature $T_{\rm N}\!\sim\!22$ K, indicating that the Eu moments are aligned parallel to the $c$ axis\cite{May2014PRB}.
To reveal the AFM arrangement of the Eu sublattice, we have measured the resonant x-ray scattering spectra near the Eu $L_{3}$ absorption edge.
At 5 K, we found the (0 0 11) reflection at $E$=6.975 keV that is forbidden in the present space group ($I4/mmm$) (inset to Fig. 1B).
Considering the evolution of the reflection intensity below $T_{\rm N}$ (Fig. 1B) and the observation of polarization rotation as well as a sharp resonance at the Eu $L_{3}$ edge (Figs. S1), it can be assigned to resonant magnetic scattering from the Eu sublattice.
Based on the analyses on several magnetic reflections (Figs. S2 and S3), we derive the most probable magnetic structure as shown in Fig. 1E.
The Eu moments order ferromagnetically in the $ab$ plane and align along the $c$ axis in the sequence of up-up-down-down, where the Bi square net intervenes between the Eu layers with magnetic moments up and down.
This magnetic arrangement can be regarded as a natural spin-valve like superstructure.
%
\par
%
Figures 1C and 1D show the temperature profiles of in-plane ($\rho_{xx}$) and interlayer ($\rho_{zz}$) resistivity for EuMnBi$_{2}$, respectively.
At 0 T below 120 K, both the $\rho_{xx}$ and $\rho_{zz}$ curves show metallic behavior down to $T_{\rm N}$, but the anisotropy is fairly large (e.g., $\rho_{zz}/\rho_{xx}\!\sim\!480$ at 50 K at 0 T).
At $T_{\rm N}$, we observed a small drop (or cusp-like anomaly) in $\rho_{xx}$ and a steep jump in $\rho_{zz}$ toward the lowest temperature.
These transport properties seem to be consistent with the antiferromagnetic order of the Eu layer; the interlayer conduction should be suppressed by the staggered Eu moments along the $c$ axis, whereas the ferromagnetic order within the plane may promote the in-plane one.
More interestingly, the application of the field parallel to the $c$ axis has a strong impact on the temperature profiles of $\rho_{xx}$ and $\rho_{zz}$ (red curves in Figs. 1C and 1D).
At 9 T, $\rho_{xx}(T)$ exhibits marked positive magnetoresistance effects that evolve with decreasing temperature down to $\sim$40 K, followed by a steep drop at $T_{\rm N}$.
On the other hand, $\rho_{zz}$ at 9 T shows minimal (longitudinal) magnetoresistance effects above $T_{\rm N}$, but shows a much larger jump on cooling at $T_{\rm N}$ than that at 0 T.
These suggest that the increase of anisotropy in resistivity below $T_{\rm N}$ is further enhanced at 9 T; in fact, the increase in $\rho_{zz}/\rho_{xx}$ with decreasing temperature from 25 K (just above $T_{\rm N}$) to 2 K exceeds 1,000\% at 9 T, whereas it is approximately 180\% at 0 T.
Judging from the temperature profile of $M/H$ at 7 T for $H||c$ in Fig. 1A (and also magnetic phase diagram in Fig. 1F), the Eu moments are oriented perpendicular to the $c$ axis in the AFM phase at 9 T, which appears to strongly suppress the interlayer conduction between the Bi layers.
We will again discuss the effect of the Eu spin flop on the resistivity based on its field profile ($vide$ $infra$).
%
\par
%
The magnetotransport properties enriched by the Eu magnetic order are further highlighted by the magnetization and resistivity measured in the magnetic field up to 55 T applied along the $c$ axis (Fig. 2).
The magnetization at 1.4 K exhibits a clear metamagnetic (spin-flop) transition at $H\!=\! H_{\rm f}$ ($\sim\!5.3$ T), corresponding to the reorientation of the Eu moments to be perpendicular to the field (Fig. 2A).
In the forced ferromagnetic state above $H_{\rm c}$ ($\sim$22 T), the magnetization is saturated close to 7 $\mu_{\rm B}$, reflecting the full moment of localized Eu 4$f$ electrons.
The temperature variation of $H_{\rm f}$ and $H_{\rm c}$ is plotted in Fig. 1F (see Fig. S4 for details), which forms typical phase diagram for an anisotropic antiferromagnet in the field parallel to the magnetization-easy axis.
%
\par
%
The interlayer resistivity is markedly dependent on the AFM states of the Eu sublattice (Fig. 2B).
Above $T_{\rm N}$ (at 27 and 50 K), $\rho_{zz}$ is almost independent of field, except for clear Shubnikov-de Haas (SdH) oscillations at 27 K.
At 1.4 K, on the other hand, $\rho_{zz}$ exhibits a large jump at $H_{\rm f}$, followed by giant SdH oscillations that reach $\Delta \rho_{\rm osc}/\rho\!\sim$50\%.
This high-$\rho_{zz}$ state is terminated at $H_{\rm c}$, above which the $\rho_{zz}$ value is substantially reduced.
The origin of such $\rho_{zz}$ enhancement (i.e., suppression of interlayer coupling) in the spin-flop phase remains as an open question at present; the interlayer charge transfer caused by electron's hopping on the local Eu moments would not change, if Eu moments were simply reoriented perpendicular to the $c$ axis while keeping the same AFM pattern.
We should note here that the Mn sublattice that antiferromagnetically orders at $\sim$315 K\cite{May2014PRB} as well as the Eu one plays a vital role on achieving the high $\rho_{zz}$ state.
As shown in inset to Fig. 2B, in fact, the $\rho_{zz}$ value at 0 T for EuZnBi$_2$ is reduced to one twenty-fifth of that for EuMnBi$_2$, although the plausible AFM order of Eu sublattice at 0 T for EuZnBi$_2$ is analogous to that in the spin-flop phase for EuMnBi$_2$ (i.e., the Eu moments are aligned in the $ab$ plane with the staggered stacking along the c axis. See Fig. S5E).
For SrMnBi$_2$, on the other hand, the $\rho_{zz}$ value at 0 T is comparable to that for EuMnBi$_2$, but shows a minimal magnetoresistance effect up to 9 T.
The magnetic order in both Eu and Mn sublattices is thus essential for enhancing $\rho_{zz}$.
As a possible model based on these facts, the magnetic order of Mn sublattice might be significantly modulated upon the Eu spin flop due to the $f$-$d$ coupling.
It is also likely that we need to take into consideration the anisotropy of Eu$^{2+}$ 4$f$ orbital induced by the crystal field splitting\cite{Laan2008PRL}, which might reduce wave function overlap with the Mn sites along the c-axis when the Eu moment and orbital rotate.
It would however be an issue for future experimental and theoretical works to reveal the detailed mechanism.
%
\par
%
Another important feature is that the $\rho_{zz}$ peak around 20 T shows a sizable hysteresis between the field-increasing and decreasing runs.
[Correspondingly, a hysteretic anomaly also manifests itself in $\rho_{xx}$ (Fig. 2C).] 
Since no clear anomaly is discerned in the magnetization curve around 20 T (Fig. S6), the Eu moments play a minor role, instead, a possible transition between the Landau levels with different spin orientation might be responsible for this hysteresis, as discussed later.
%
\par
%
The in-plane resistivity exhibits a large positive (transverse) magnetoresistance effect, irrespective of the Eu magnetic order (Fig. 2C).
At 50 K, the $\rho_{xx}(H)$ profile is strikingly $H$-linear without saturation up to 35 T, resulting in the magnetoresistance ratio of $\rho(H\!=\!35\ {\rm T})/\rho(0)\!\sim$ 2,000\%.
Such large $H$-linear magnetoresistance is occasionally observed in Dirac semimetals\cite{Liang2014NatMat, He2014PRL, Narayanan2015PRL, Kushwaha2015APLMat, Novak2015PRB}.
At lower temperatures, the SdH oscillations are superimposed; at 1.4 K, in particular, the magnitude of oscillation is largely enhanced in the spin-flop AFM phase between $H_{\rm f}$ and $H_{\rm c}$, similarly to $\rho_{zz}$.
The enhanced SdH oscillations in the spin-flop phase are also noticeable for the Hall resistivity $\rho_{yx}$ (Fig. 2D), which show plateau-like structures at 1.4 K.
In the following, we will analyze the details of $\rho_{yx}$ plateaus in terms of the multilayer QHE in the stacking 2D Bi layers.
%
\par
%
In Fig. 3A, we plot the inverse of $\rho_{yx}$ at 1.4 K (spin-flop phase) as a function of $B_{\rm F}/B$, where $B_{\rm F}$ is the frequency of SdH oscillation and $B$ is the magnetic induction.
$B_{\rm F}/B$ is the normalized filling factor [corresponding to $(n+\frac{1}{2}-\gamma)$ in Eq. (\ref{eq:Ryx})]\cite{Lukyanchuk2006PRL}, which is employed to compare the samples with different $B_{\rm F}$ (Table 1).
The inverse of $\rho_{yx}$ also exhibits clear plateaus at regular intervals of $B_{\rm F}/B$, the positions of which nicely correspond to deep minima in $\rho_{xx}$ (Fig. 3B) and pronounced peaks in $\rho_{zz}$ (Fig. 3C).
All these features signify the multilayer QHE, as previously observed for the GaAs/AlGaAs superlattice\cite{Druis1998PRL, Kuraguchi2000PhysicaE}.
Although the $\rho_{xx}$ minima do not reach zero, $\omega_{\rm c}\tau$ estimated from $\rho_{yx}/\rho_{xx}$ is much larger than unity (e.g. $\sim$5 at around $B_{\rm F}/B\!=\!1.5$, see Fig. S7A), where $\omega_{\rm c}$ the cyclotron frequency and $\tau$ the scattering time.
What is prominent in the present compound is that the values of $1/\rho_{yx}$ are quantized to half-integer multiples, when scaled by $1/\rho^0_{yx}$, the step size of successive plateaus (see Fig. S7B for definition).
Given the conventional view of QHE, this quantization of $\rho^0_{yx}/\rho_{yx}$ leads to the normalized filling factor of $n\!+\!\frac{1}{2}$, where $n$ is a non-negative integer.
This is consistent with the plateaus occurring at half-integer multiples of $B_{\rm F}/B$ (vertical dotted lines in Fig. 3A, where a small shift corresponds to the phase factor as explained below).
Such a half-integer (normalized) filling factor is known to stem from the nontrivial $\pi$ Berry's phase of Dirac fermions\cite{Novoselov2005Nature,Zhang2005Nature}, which in two dimensions leads to the Hall resistance quantized as follows:\cite{Zheng2002PRB, Gusynin2005PRL}
\begin{equation}
 \frac{1}{R_{yx}}=\pm s\left(n+\frac{1}{2}-\gamma\right)\frac{e^2}{h},
\label{eq:Ryx}
\end{equation}
where $s$ the spin and valley degeneracy factor and $\gamma$ the phase factor expressed as $\gamma=\frac{1}{2}-\frac{\phi_B}{2\pi}$ with $\phi_B$ the Berry's phase\cite{Mikitik1999PRL}.
The observed half-integer filling factor thus corresponds to $\gamma\!\sim\!0$, i.e., nontrivial $\pi$ Berry's phase in the present QHE.
Following standard analyses on the SdH oscillations using fan diagram, furthermore, we plot the values of $1/B$ at the $\rho_{xx}$ minima (or $\rho_{zz}$ maxima) against half integers (inset to Fig. 3).
Based on a semiclassical expression of oscillating part in $\rho_{xx}$\cite{Zhang2005Nature,Lukyanchuk2006PRL}, $\Delta\rho_{xx}\!(\propto\!-\Delta\rho_{zz})\!\propto\!\cos\left[2\pi(B_{\rm F}/B-\gamma+\delta)\right]$, a linear fitting yields $\gamma\!-\!\delta$ close to 0 ($\sim\!-0.1$) for all the samples (Table 1), where a phase shift $\delta$ is determined by the dimensionality of the Fermi surface, varying from 0 (for 2D) to $\pm 1/8$ (for 3D).
Since the value of $\delta$ tends to be negligibly small for quasi 2D Fermi surfaces even in bulk materials\cite{Lukyanchuk2006PRL,Murakawa2013Science}, the fitted results indicate $\gamma\!\sim\!-0.1$, which again verifies the non-zero Berry's phase in this compound.
%
\par
%
The quantization of $\rho^0_{yx}/\rho_{yx}$ to half-integer multiples is well reproduced for two samples (\#1 and \#2 in Fig. 3A).
The thickness of sample \#2 is $\sim$60\% of that of \#1.
Nevertheless, their difference in $\rho^0_{yx}$ is only $\sim$10\%.
This fact ensures that the observed Hall plateaus are of bulk origin, which should be attributed to the parallel transport of the 2D Bi layers stacking along the $c$ axis, as is the case for multilayer quantum Hall systems, including semiconductor superlattice\cite{Druis1998PRL,Stormer1986PRL}, Bechgaard salts\cite{Cooper1989PRL, Hannahs1989PRL}, Mo$_4$O$_{11}$\cite{Hill1998PRB, Sasaki1999SSC}, and Bi$_2$Se$_3$\cite{Cao2012PRL}.
The inverse Hall resistivity is hence expressed as $1/\rho_{yx}\!=\!Z^{*}/R_{yx}$, where $Z^*\!=\!\frac{1}{c/2}\!\sim\! 8.9\times 10^6$ (cm$^{-1}$) is the number of the Bi layer per unit thickness and $c$ the $c$-axis length.
This gives the step size between the successive $1/\rho_{yx}$ plateaus as $1/\rho^0_{yx}\!=\!sZ^{*}(e^2/h)$, from which we have estimated the degeneracy factor $s$ to be $\sim$5-6 as shown in Table 1 (see also Fig. S7B and the related discussions).
Provided that there exist four valleys in EuMnBi$_2$\cite{Borisenko2015arxiv} as is the case of SrMnBi$_2$\cite{Park2011PRL,Lee2013PRB}, $s$ should be 8 (including double spin degeneracy).
Even having taken account of errors in sample thickness ($\pm$10-20\%), the $s$ value of 8 is somewhat larger than the estimated one, which may be attributable to the inhomogeneous transport arising from dead layers and/or the imperfect contacts.
%
\par
%
From the SdH frequencies in the spin-flop phase, we are capable of estimating the 2D carrier density per Bi layer at 1.4 K to be $n_{\rm 2D}\!=\!seB_{\rm F}/h\!\sim\!4.9\!\times\!10^{12}$ cm$^{-2}$ assuming $s$=8, which results in three dimensional density $n_{\rm 3D}\!=\!n_{\rm 2D}Z^{*}\!\sim\!4.4\times10^{19}$ cm$^{-3}$ (sample \#1).
This is comparable to that estimated from $\rho_{yx}$ at $\sim$20 T; $n_{\rm H}\!=\!B/e\rho_{yx}\!\sim\!2.9\pm0.2\!\times\!10^{19}$ cm$^{-3}$, where errors arise from the oscillatory component.
From the value of residual resistivity $\rho_0$, we have obtained the mobility $\mu\!=\!n_{\rm 3D}/e\rho_0\!\sim$14,000 cm$^2$/Vs at $\sim$2 K, which attains a markedly high value despite the transport coupled with the Eu magnetic order.
%
\par
%
As shown in Fig. 3B, the $N$=2 Landau level clearly splits into two peaks in the second derivative of resistivity $-d^2\rho_{xx}/dB^2$, while the splitting for $N$=3 is barely discernible.
This Landau level splitting is likely to be more pronounced for $N$=1 (at higher fields), supposedly forming a dip structure in $\rho_{xx}$ as well as $-d^2\rho_{xx}/dB^2$ around $B_{\rm F}/B$=1.
Unfortunately, only one of the split Landau levels is accessible for $N$=1, since the spin-flop phase is terminated at $H_{\rm c}$ (a spiky peak in $-d^2\rho_{xx}/dB^2$, see also Fig. 4).
With further decreasing temperature down to 50 mK, another Landau level splitting appears to evolve (thick arrow in Fig. 4).
Although the origin of these splittings is unclear at present, it should be relevant to the spin and valley degrees of freedom, as is often the case in the conventional QHE in semiconductor heterostructures\cite{Ando1982RMP}.
It is surprising that such lifting of spin and valley degeneracy is clearly observed at moderately high fields ($\sim20$ T) even in the bulk system.
This may be indicative of a large Land\'{e} $g$ factor and/or strong electron correlations, characteristic of Dirac fermions formed on the Bi layer\cite{Jo2014PRL,Zhu2011NatPhys,Li2008Science}.
%
\par
%
We finally mention about the hysteretic anomalies in $\rho_{xx}$ and $\rho_{zz}$ around 20 T (Fig. 2).
It should be noted here that similar hysteretic phenomena of resistivity have been discovered in many 2D electron gas systems both in the regimes of the integer\cite{Piazza1999Nature,Poortere2000Science} and fractional QHE\cite{Cho1998PRL,Eom2000Science}.
Their physical origin is considered to be relevant to the crossing of Landau levels for electrons (or for composite fermions in the fractional QHE) with different spin polarization\cite{Jungwirth1998PRL}, where magnetic domains are likely to form.
In the present compound, since the resistivity shows substantial hysteresis near the transition between the split Landau levels (in the $N$=1 state as shown in Fig. 4), it might originate from the dissipative conduction along such domain walls.
While detailed discussions about its mechanism are beyond the scope of the present study, the observed distinct hysteresis may suggest the possible importance of the spin-polarization of Landau level for Dirac fermions.
%
\par
%



%
\par
%
We have here presented a dramatic tuning of magnitude in interlayer conduction of quasi 2D Dirac fermions, utilizing the AFM order of Eu moments.
In addition to the staggered moment alignment along the $c$ axis, the field-induced flop of the Eu moment direction appears to further reduce the interlayer coupling and hence confine the Dirac fermions within the constituent 2D Bi layer well enough to quantize the Hall conductivity in a bulk form\cite{Chalker1995PRL}.
Such a magnetically-active Dirac fermion system would form a promising class of spintronic materials with very high mobility.
\section*{Materials and Methods}
%
Single crystals of EuMnBi$_2$, SrMnBi$_2$, and EuZnBi$_2$ were grown by a Bi self-flux method.
For EuMnBi$_2$, high purity ingots of Eu (99.9$\%$), Mn (99.9$\%$), Bi (99.999$\%$) were mixed in the ratio of Eu:Mn:Bi = 1:1:9 and put into an alumina crucible in an argon-filled glove box.
For SrMnBi$_2$, the ratio of the mixture was Sr:Mn:Bi=1:1:9, while it was Eu:Zn:Bi=1:5:10 for EuZnBi$_2$.
The crucible was sealed in an evacuated quartz tube and heated at 1000$^\circ\mathrm{C}$ for 10 h, followed by slow cooling to 400$^\circ\mathrm{C}$ at the rate of $\sim2^\circ\mathrm{C/h}$, where the excess Bi flux was decanted using a centrifuge\cite{Canfield2001JCG}.
Plate-like single crystals with a typical size of $\sim 5\times5\times1\ \mathrm{mm^3}$ were obtained.
The powder x-ray diffraction profiles at room temperature indicate that the crystal structure is tetragonal ($I4/mmm$) for all the materials (see Figs. S8A-S8C).
From Le Bail fitting of the measured profiles, the lattice constants are estimated to be $a$=4.5416(4) \AA \ and $c$=22.526(2) \AA, $a$=4.5609(4) \AA \ and $c$=23.104(2)\AA, and $a$=4.6170(3) \AA \ and $c$=21.354(2) \AA \ for EuMnBi$_2$, SrMnBi$_2$ and EuZnBi$_2$, respectively.
%
\par
%
At low fields, magnetization (up to 7 T) and resistivity (up to 14 T) were measured down to 1.9 K using Magnetic Property Measurement System (Quantum Design) and Physical Properties Measurement System (Quantum Design), respectively.
In-plane resistivity $\rho_{xx}$ and Hall resistivity $\rho_{yx}$ were measured by a conventional 5-terminal method with electrodes formed by room-temperature curing silver paste (Fig. S8D).
The typical sample dimension is $\sim2.0$ mm (length) $\times0.5$ mm (width) $\times0.1$ mm (thickness).
The voltage terminals were needed to cover the whole thickness of the sample side to avoid the admixture of the interlayer resistance.
Interlayer resistivity $\rho_{zz}$ was measured by 4-terminal method on bar-shaped samples with a typical dimension of $\sim1.5 \ \mathrm{mm}$ in length along the $c$ axis and $\sim 0.4\times0.4\ \mathrm{mm^2}$ in cross section  (Fig. S8E).
Current terminals were formed so as to completely short out the in-plane current.
The magnetization and resistivity up to 55 T were measured using the non-destructive pulsed magnet with a pulse duration of $36\ \mathrm{msec}$ at the International Mega-Gauss Science Laboratory at the Institute for Solid State Physics.
The measurement temperature range was 1.9-150 K.
The magnetization was measured by the induction method, using coaxial pickup coils.
The resistivity ($\rho_{xx}$, $\rho_{yx}$, and $\rho_{zz}$) was measured by a lock-in technique at 100 kHz with the ac excitation of 1-10 mA.
The resistivity measurement up to 28 T at $\sim 50\ \mathrm{mK}$ was performed with a lock-in amplifier at 17 Hz with the ac excitation of $100\ \mathrm{\mu A}$ by using a dilution refrigerator embedded in the cryogen- free hybrid magnet at High Field Laboratory for Superconducting Materials in Institute of Materials Research, Tohoku University\cite{HighField_Tohoku}.
%
\par
%
Resonant x-ray magnetic scattering measurements were performed at BL-3A, Photon Factory, KEK, Japan, by utilizing the horizontally polarized x-ray in resonance with Eu $L_3$ absorption edge ($\sim$6.975 keV).
We employed a relatively thick sample with the (0 0 1) and (1 0 $l$) natural facets ($l\!\sim$1-2) with the dimension of $\sim$3$\times$2$\times$1.5 mm$^{3}$.
The (0 0 $L$) and (1 0 $L$) reflections were measured on the (0 0 1) natural facet (inset to Fig. S1B), by attaching the sample to the cold finger of a He closed-cycle refrigerator on a four-circle diffractometer (5 K-300 K).
The (4 0 1) and (3 0 0) reflections were measured on the (1 0 $l$) natural facet (inset to Fig. S1C), using a liquid helium flow type cryostat on a two-circle diffractometer (5 K-40 K).
For selected magnetic reflections, we performed polarization rotation measurements, where the polarization of scattered x-rays was analyzed by utilizing a Cu(110) single crystal.
Unless otherwise stated, the scattered x-rays were detected without analyzing polarization and hence includes both the $\sigma^\prime$- and $\pi^\prime$-polarization components.
For all the measurements, we used a silicon drift detector (SDD).
\section*{Supplementary Materials}
Supplementary Material accompanies this paper at {\small {\tt http://www.scienceadvances.org/}}.\\
Fig. S1. Resonant x-ray magnetic scattering for EuMnBi$_2$ near the Eu $L_3$ absorption edge.\\
Fig. S2. Extinction rules and candidates of magnetic structure of Eu sublattice\\
Fig. S3. Determination of magnetic structure of Eu sublattice.\\
Fig. S4. Detailed magnetic properties for EuMnBi$_2$.\\
Fig. S5. Magnetic properties for EuZnBi$_2$\\
Fig. S6. Magnetization and transport features around 20 T\\
Fig. S7. Hall angle and step size between the consecutive $1/\rho_{yx}$ plateaus\\
Fig. S8. Powder x-ray diffraction profile for each compound and geometry of the samples and electrodes\\
References \cite{XRMS_EuRh2As2,XRMS_Hannon}\\
%
%
%
\bibliographystyle{ScienceAdvances}

\noindent \textbf{Acknowledgements:} 
%
The authors are grateful to Y. Motome, S. Hayami, J. Fujioka, R. Yoshimi, K. Akiba, and T. Osada. The measurements in KEK were performed under the approval of the Photon Factory Program Advisory Committee (Proposal No. 2015S2-007, 2013G733, and 2012S2-005). Low temperature measurement in dilution fridge was performed at the High Field Laboratory for Superconducting Materials, Institute for Materials Research, Tohoku University.\\
\noindent \textbf{Funding:}  This study was partly supported by the Japan Society for the Promotion of Science (JSPS) Grants-in-Aid for Exploratory Research No. 15K13332, Mizuho Foundation for the Promotion of Sciences, Nippon Sheet Glass Foundation for Materials Science, Asahi Glass Foundation, and JST PRESTO program (Hyper-nano-space design toward Innovative Functionality).\\
\noindent \textbf{Author Contributions} H.S. conceived the project. H.S. and S.I. designed and guided the experiments. H.M. and H.S. performed single crystal growth. H.M., H.S., M.T., and A.M. measured transport and magnetic properties at high fields. H.M., H.S., Y.Y., H.N., Y.M., and T.A. performed resonant x-ray magnetic scattering experiments. J.S., A.T., S.N., and S.A. measured the resistivity using a dilution fridge. A.T., T.A., and Y.T. contributed to the discussion of the results. H.S., S.I., Y.T., and H.M. wrote manuscript with contributions from all authors. \\
\noindent \textbf{Competing Interests} The authors declare that they have no competing interests.\\
\noindent \textbf{Data and materials availability:} All data are presented in the article and Supplementary Materials. Please direct all inquiries to the corresponding author.
%
%
\clearpage
%
\begin{table}
\caption{Parameters related to the SdH oscillations and quantized Hall plateaus in the spin-flop phase (at 1.4 K and 5.3-22 T). $B_{\rm F}$ and $\gamma$ are the results of linear fit to the Landau fan plot.}
\vspace*{.5cm}
\hspace*{-1cm}
\begin{tabular}{c|c|c|c|c|c|c}%
Sample \#&$\rho_{ij}$&$B_{\rm F}$&$\gamma$&Sample thickness &$\rho^0_{yx}$ &$s$\\
         &          &  (T)       &(phase factor)&($\mu$m)   &($\mu\Omega$cm) &(degeneracy factor)\\\noalign{\hrule height1.5pt}
    1    &$\rho_{xx}$, $\rho_{yx}$& $26.1(2)$    &$-0.12(4)$         &130                    & 525            &5.5\\\hline
    2    &$\rho_{xx}$, $\rho_{yx}$& $23.1(1)$    &$-0.12(2)$         &78                   & 578            &5.0\\\hline
    3    &$\rho_{zz}$             & $19.5(1)$    &$-0.08(2)$         &-                       & -            &-\\\hline
\end{tabular}
\end{table}
%
\clearpage

%
\noindent {\bf Fig. 1.} {\bf Transport coupled with the magnetic order of Eu sublattice.} {\bf (A-D)} Temperature dependence of magnetic and transport properties near the antiferromagnetic transition temperature ($T_N$) for EuMnBi$_2$. {\bf (A)} Magnetic susceptibility $M/H$ for the field parallel to the $c$ axis ($H||c$) at 0.1 T (blue) and 7 T (red). Open symbols are the data for the field perpendicular to the $c$ axis ($H\! \perp \!c$) at 0.1 T. {\bf (B)} Intensity of resonant magnetic reflection (0 0 11) at 6.975 keV at 0 T. The inset shows the profile of the (0 0 11) reflection along [001] at 6.975 keV (resonant) and 7.00 keV (nonresonant). {\bf (C)} In-plane resistivity $\rho_{xx}$ and {\bf (D)} interlayer resistivity $\rho_{zz}$ at 0 and 9 T ($H||c$). Schematic sample configuration for the resistivity measurement is shown in each panel. {\bf (E)} Schematic illustration of the plausible magnetic structure for EuMnBi$_2$ at zero field, together with the formal valence of each ion. The arrangement of the Mn sublattice is assumed to be the same as in SrMnBi$_2$\cite{Guo2014PRB}. {\bf (F)} Magnetic phase diagram for the Eu sublattice as functions of field ($H||c$) and temperature. PM and AFM denote the paramagnetic and antiferromagnetic states, respectively. $H_{\rm f}$ and $H_{\rm c}$ correspond to the transition fields to the spin-flop AFM and PM (forced ferromagnetic) phases, respectively. Black arrows are schematic illustration of the Eu moments sandwiching the Bi$^{-}$ layer. Note the Mn sublattice orders at $\sim$315 K ($>T_{\rm N}$).
\\
\\
%
\noindent {\bf Fig. 2.} {\bf Magnetic field dependence of magnetic and transport properties at high fields.} {\bf(A)} $M$, {\bf(B)} $\rho_{zz}$ (sample \#3), {\bf(C)} $\rho_{xx}$, and {\bf(D)} Hall resistivity $\rho_{yx}$ (sample \#1) for EuMnBi$_2$ at selected temperatures for the field parallel to the $c$ axis up to $\sim$55 T. Schematic illustration of the Eu$^{2+}$ moments adjacent to the Bi layer for $H\!<\!H_{\rm f}$, $H_{\rm f}\!<\!H\!<\!H_{\rm c}$, and $H_{\rm c}\!<\!H$ is presented in {\bf(A)}. The inset to {\bf(B)} shows the field profile of $\rho_{zz}$ (below 9 T) for EuMnBi$_2$, EuZnBi$_2$, and SrMnBi$_2$.
\\
\\
%
\noindent {\bf Fig. 3.} {\bf Quantized Hall plateaus and SdH oscillations.} {\bf(A)} Normalized inverse Hall resistivity $\rho^0_{yx}/\rho_{yx}$ versus $B_{\rm F}/B$ measured at 1.4 K for samples \#1 and \#2, where $B_{F}$ is the frequency of Shubnikov-de Haas (SdH) oscillation and $B=\mu_0(H+M)$ the magnetic induction. $1/\rho^0_{yx}$ is the step size between the consecutive plateaus in $1/\rho_{yx}$ (see Fig. S7B). {\bf (B)} $\rho_{xx}$, second field derivative $-d^2 \rho_{xx}/dB^2$ for sample \#1 and {\bf (C)} $\rho_{zz}$ for sample \#3 versus $B_{\rm F}/B$ measured at 1.4 K. Vertical dotted lines denote half-integer multiples shifted by $-\gamma$, where $\gamma\!\sim\!-0.1$ is a phase factor estimated from the fan plot. (Inset) Landau fan plot ($1/B$ versus $N$) for \#1, \#2 and \#3. The slope and intercept with the $N$ axis give $B_{\rm F}$ and $\gamma\!-\!\delta$, respectively (Table 1). A phase shift $\delta$ should be negligibly small for a quasi 2D Fermi surface, as discussed in the main text.
\\
\\
%
\noindent {\bf Fig. 4.} {\bf Hysteresis and split of the Landau level.} $\rho_{xx}$ as a function of $1/B$ at 50 mK and 1.4 K. The curve at 50 mK is shifted downward for clarity. The arrow denotes the $\rho_{xx}$ valley noticeable at 50 mK. Long solid and dashed lines indicate the integer and half-integer multiple of $B_{\rm F}/B-\gamma$, respectively. Short solid line denotes the position corresponding to the field $H\!=\!H_{\rm c}$.\\
\\
%
\begin{figure}
\begin{center}
\includegraphics[width=\linewidth]{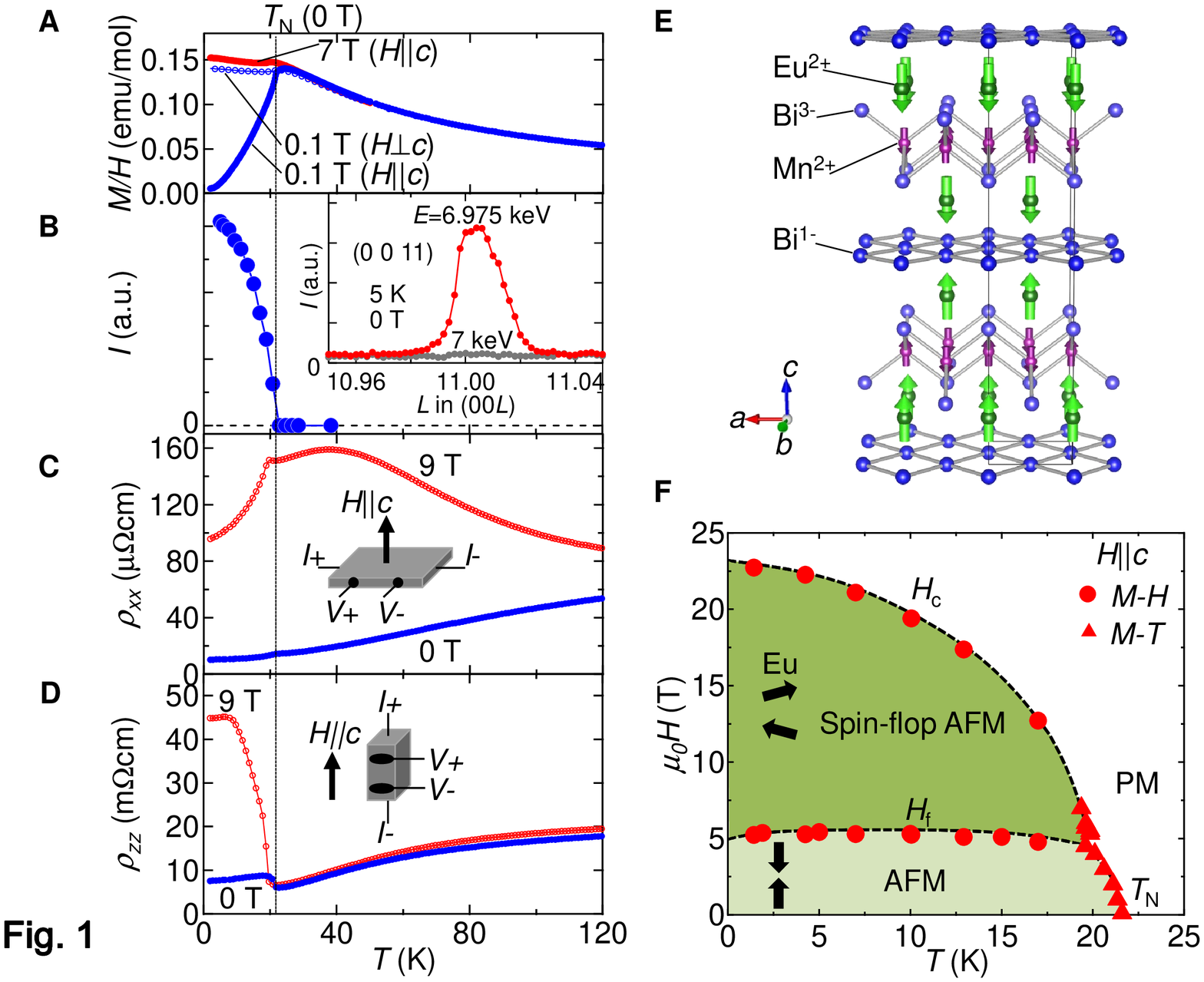}
\end{center}
\end{figure}
%
\begin{figure}
\begin{center}
\includegraphics[width=0.85\linewidth]{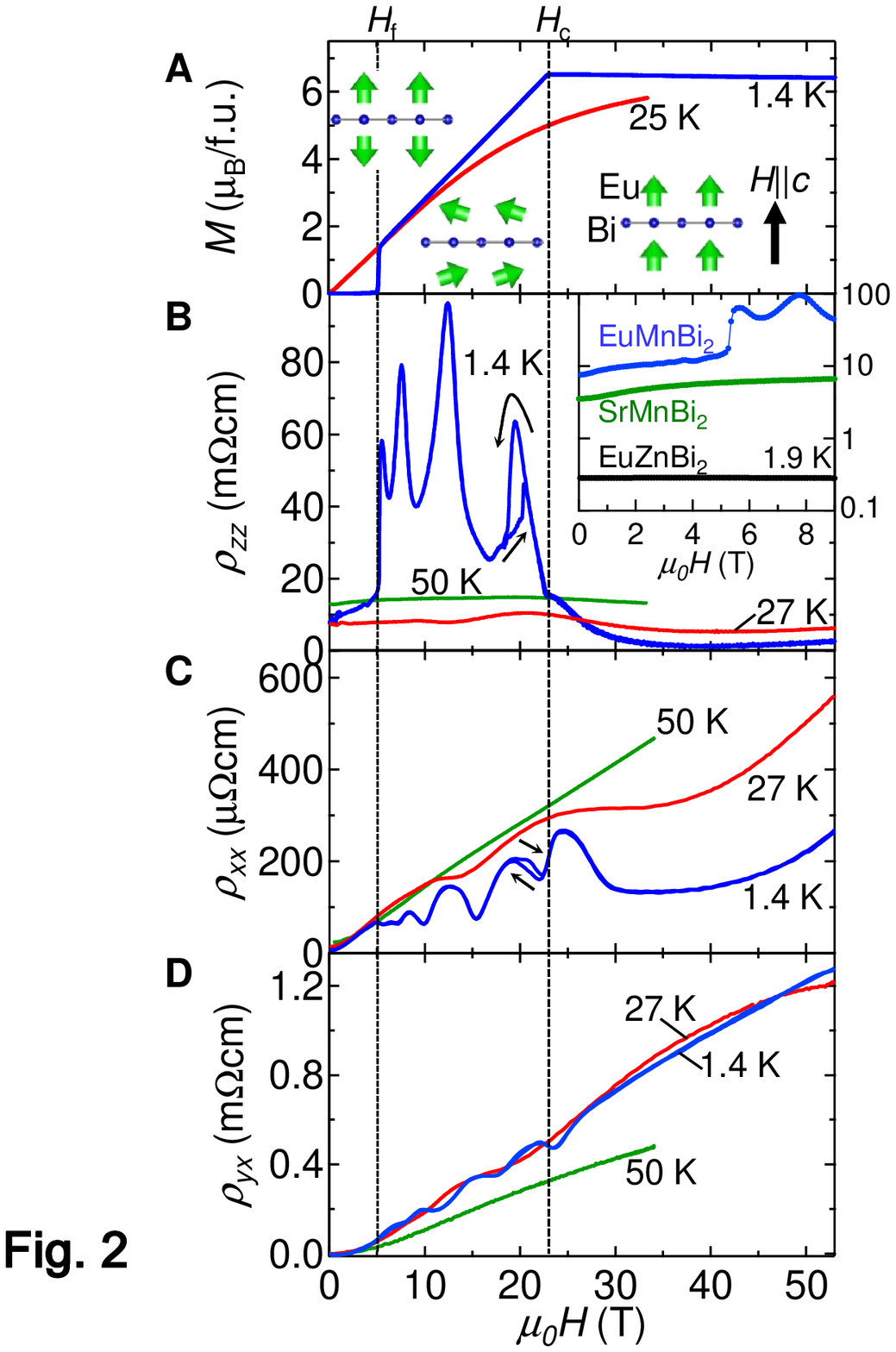}
\end{center}
\end{figure}
%
\begin{figure}
\begin{center}
\includegraphics[width=0.9\linewidth]{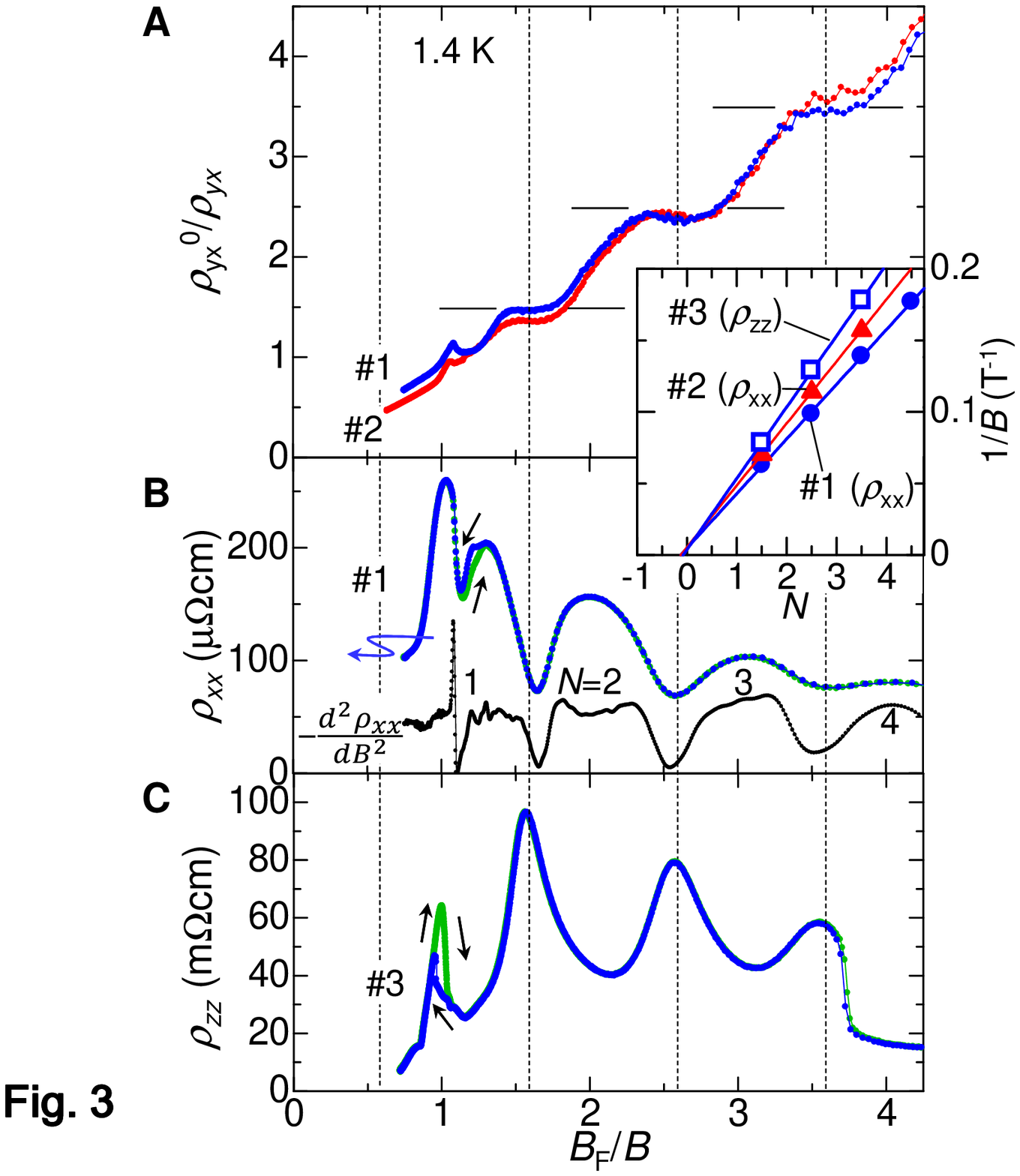}
\end{center}
\end{figure}
%
\begin{figure}
\begin{center}
\includegraphics[width=0.9\linewidth]{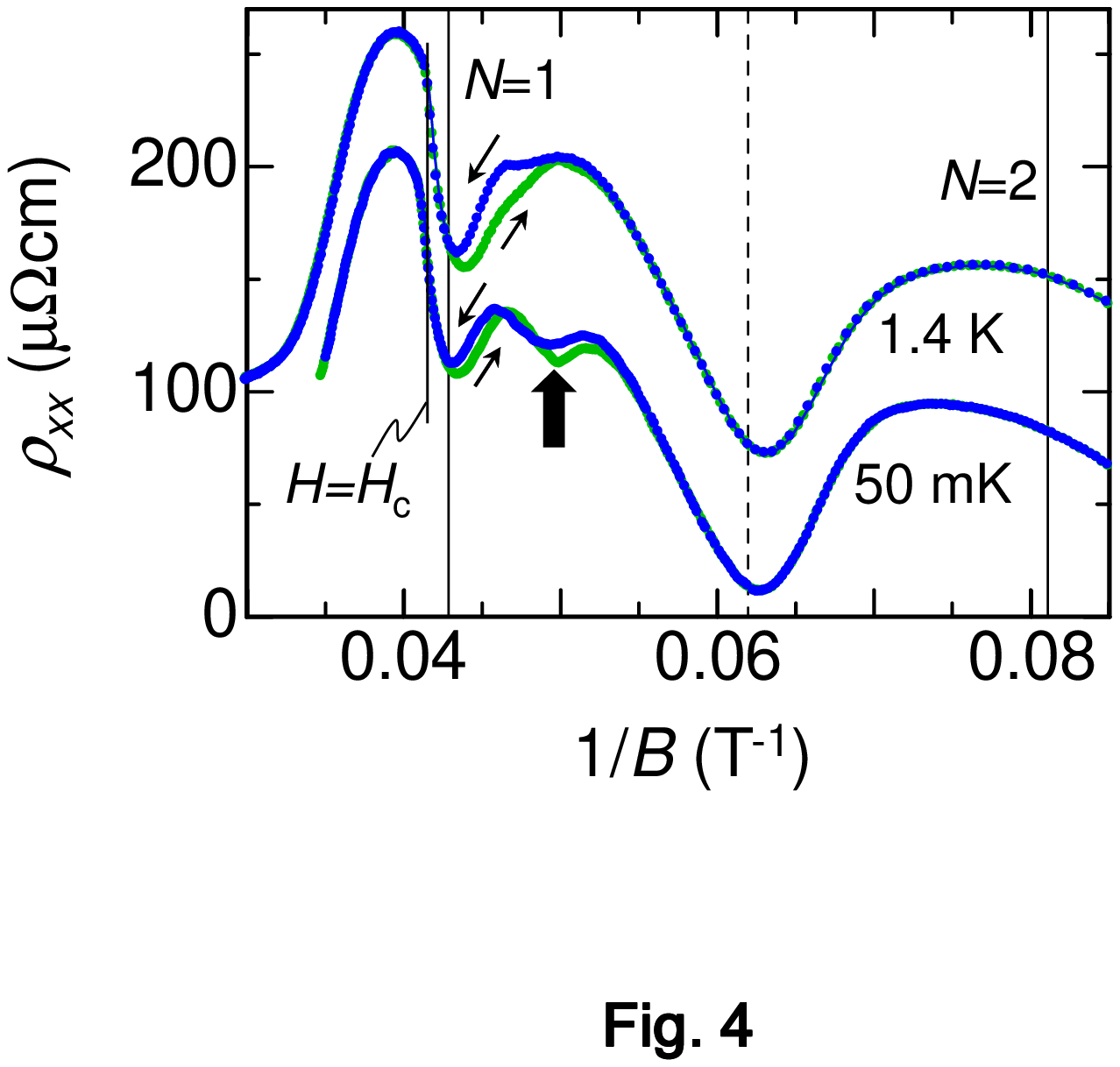}
\end{center}
\end{figure}
%
\end{document}